# Monolayer Two-dimensional Materials Database (ML2DDB) and Applications

Zhongwei Liu[a, b, #], Zhimin Zhang[c, #], Xuwei Liu[c, #], Mingjia Yao[b], Xin He[a], Yuanhui Sun[b, *], Xin Chen[b, *], Lijun Zhang[a, b, *]

[a]*State Key Laboratory of Integrated Optoelectronics, Key Laboratory of Automobile Materials of MOE and College of Materials Science and Engineering, Jilin University, Changchun 130012, China*
[b]*Suzhou Laboratory, Suzhou, 215123, China*
[c]*Baidu Inc., Beijing, P.R. China.*
[#]These authors contributed equally to this work.
E-mail: sunyh@szlab.ac.cn; chenx01@szlab.ac.cn; lijun_zhang@jlu.edu.cn

# Abstract

The discovery of two-dimensional (2D) materials with tailored properties is critical to meet the increasing demands of high-performance applications across flexible electronics, optoelectronics, catalysis, and energy storage. However, current 2D material databases are constrained by limited scale and compositional diversity. In this study, we introduce a scalable active learning workflow that integrates deep neural networks with density functional theory (DFT) calculations to efficiently explore a vast set of candidate structures. These structures are generated through physics-informed elemental substitution strategies, enabling broad and systematic discovery of stable 2D materials. Through six iterative screening cycles, we established the creation of the Monolayer 2D Materials Database (ML2DDB), which contains 242,546 DFT-validated stable structures—an order-of-magnitude increase over the largest known 2D materials databases. In particular, the number of ternary and quaternary compounds showed the most significant increase. Combining this database with a generative diffusion model, we demonstrated effective structure generation under specified chemistry and symmetry constraints. This work accomplished an organically interconnected loop of 2D material data expansion and application, which provides a new paradigm for the discovery of new materials.

# Introduction

The exploration and utilization of novel materials are increasingly recognized as key drivers for advancing cutting-edge technologies and upgrading industrial systems. Two-dimensional (2D) materials, characterized by their atomic-scale thickness, quantum confinement effects [1], and high specific surface area [2], hold great promise in areas such as flexible electronics [3], optoelectronics [4], catalysis [5], and energy storage [6]. However, with the growing complexity and specificity of performance requirements (particularly in thermal, mechanical, and optical domains [7–9]), existing 2D materials are often fall short of meeting practical performance demands. To address these challenges, it is essential to systematically expand the library of 2D materials and thoroughly explore their multidimensional properties, thereby accelerating the discovery of candidates tailored to specific application needs [10,11].

With the rapid rise of data-driven materials science, leveraging existing databases has become a powerful strategy for accelerating materials design and discovery [12–14]. Large-scale databases now encompass hundreds of thousands to millions of inorganic and organic structures, such as Inorganic Crystal Structure Database [15], Materials Project [16], Open Quantum Materials Database (OQMD) [17], and Quantum Mechanics 9 (QM9) [18], enabled by curated aggregation and elemental substitution techniques. The GNoME model, developed by Google DeepMind, integrates deep learning with density functional theory (DFT) calculation to produce ~2.2 million inorganic crystal structures, achieving an energy prediction mean absolute error (MAE) of just 21 meV/atom [19]. In the domain of 2D materials, new candidates are typically generated by applying techniques such as the topological scaling algorithm [20] or relative lattice-constant error analysis [21], followed by physics-informed elemental substitution. To date, the largest DFT-validated 2D materials database is Computational 2D Materials Database (C2DB), which includes more than 16,000 2D materials [22]. Nonetheless, the size of 2D materials datasets remains one to two orders of magnitude smaller than those of their 3D counterparts (OQMD contains a million structures). This highlights the urgent need to establish a closed-loop, data-driven framework capable of systematically predicting and screening the vast material space defined by nearly one hundred elements and diverse stoichiometries. Such a framework would not only enable the efficient identification of thermodynamically stable 2D materials, but also substantially enrich the diversity of candidate materials tailored to a wide range of technological applications.

In this work, we developed an active learning framework that integrates deep neural networks with DFT calculations, culminating in the creation of the Monolayer 2D Materials Database (ML2DDB). This database contains over 242,546 DFT-validated stable monolayer structures, representing an order-of-magnitude increase over the largest known 2D materials database. Notably, the dataset exhibits high compositional diversity, with elemental coverage extending across nearly the entire periodic table. The number of ternary compounds increased by 1100%, while quaternary compounds saw a 960% increase. Our machine learning interatomic potentials (MLIP) were trained on a dataset comprising 1,863,788 structure–energy–force mappings derived from 392,319 2D materials. The resulting MLIP show high prediction accuracy, reaching a success rate of 92.36%. In pursuit of further expansion of the material design space, we constructed the conditionally constrained diffusion generation model, a framework that facilitates the generation of novel structures defined by specified elemental components or properties. This model empowers us to idenitfy 2D materials that are both stable and capable of meeting target property requirements with enhanced efficiency. This work not only expanded the design space of monolayer 2D materials, but also established a closed-loop framework for conditionally guided structural exploration and generation.

## Results and discussion

### Expansion of 2D materials dataset and conditional diffusion-based structure generation

Combining data expansion with conditional diffusion-based structure generation can effectively improve the efficiency of research in designing materials well aligned with the target requirements. To enable large-scale generation and screening of candidate crystal structures, we developed a closed-loop active learning framework (Figure 1), consisting of four key modules: 2D materials data collection, structure expansion via physics-guided element substitution, MLIP-accelerated structure screening, and DFT-based validation. After multiple iterative processes, the framework progressively expands the 2D materials dataset while enhancing the model's screening capabilities. Based on the expanded database of thermodynamically stable materials, we have further carried out conditional diffusion-based structure generation. By incorporating crystal graphs, monolayer thickness, and target properties into model training, the generation of 2D materials under target property constraints can be effectively enabled. We show here the design of materials with given elemental components and space group properties. This design flow can be followed for more properties such as carrier mobility, band gap, and magnetic properties.

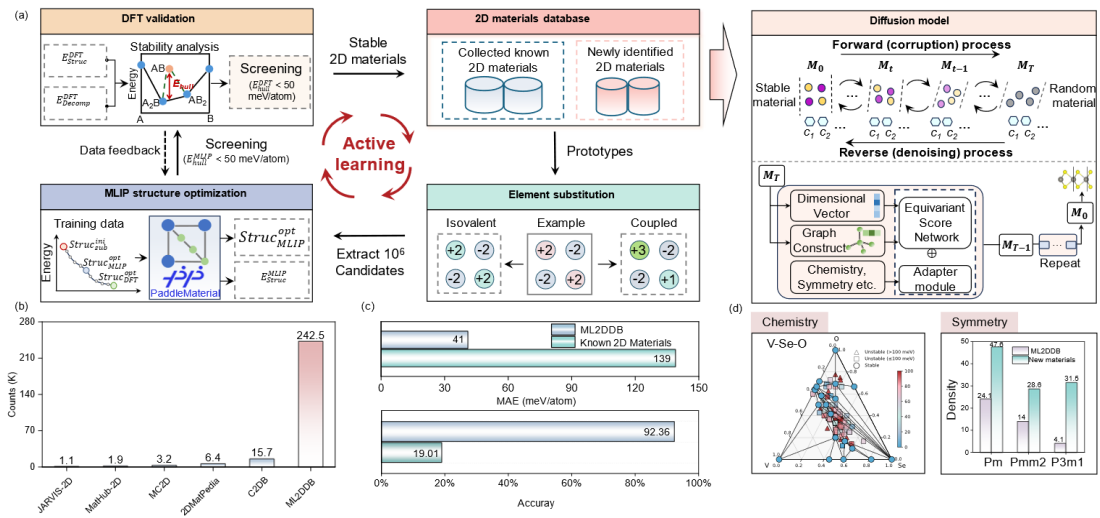

**Figure 1: Schematic diagram of active learning framework and results summary.** (a) The framework automates material screening and application through five modules: 2D materials data collection, structure expansion via physics-guided element substitution, MLIP-accelerated structure screening, DFT-based validation, and conditional diffusion-based structure generation. (b) The active learning framework discovers 242.5 thousand novel and stable materials ($E_{hull}^{DFT} < 50$ meV/atom), representing more than an order-of-magnitude increase in the number of unique structures. (c) According to the dataset expansion, the energy prediction MAE of MLIP for 2D materials reduces from 139 to 41 meV/atom, and the prediction accuracy for stable 2D materials improves from 19.01% to 92.36%. (d) Diffusion-based structure generation for given chemsitry (elemental components) and symmetry (space groups).

The workflow begins with prototype identification from known 2D materials, followed by a physics-informed element substitution strategy to generate a candidate phase space comprising over hundreds of millions hypothetical structures. In each active learning round, one million structures are randomly sampled and optimized using a trained MLIP [23,24]. Those with MLIP-predicted convex hull energies ($E_{hull}^{MLIP}$) below 50 meV/atom [25,26] are selected for DFT validation. Structures satisfying DFT-verified convex hull energies ($E_{hull}^{DFT}$) below 50 meV/atom are then incorporated into the growing materials database. Results from each DFT round are fed back into the MLIP training process, continuously improving the accuracy and efficiency of the screening pipeline. After five active learning iterations, the resulting MLIP for 2D materials achieved a MAE of 41 meV/atom and reached a prediction accuracy of 92.36% for identifying stable structures with $E_{hull}^{DFT}$ < 50 meV/atom. Depending on the DFT-validated ML2DDB, we have trained an equivariant score network diffusion model [27,28] that learns the joint distribution of atomic coordinates, lattice parameters, and chemical compositions. By using existing high-precision MLIP models for rapid structure optimization, we can obtain phase diagrams for given 2D materials system and efficient generation with specific space group constraints.

**Structure expansion via physics-guided element substitution**

A total of 21,684 unique 2D materials were compiled by aggregating publicly available databases and structural deduplication. The available databases contain JARVIS-2D [29], MatHub-2d [30], MC2D [31], 2DMatPedia [32], and C2DB [22]. From these materials, 3,512 distinct structural prototypes were identified and subsequently used for candidate structure generation.

To comprehensively explore the structural phase space of stable 2D materials, we performed elemental substitution on 3,512 structural prototypes using ionic similarity probabilities [33]. Taking the *P6$_3$/mmc* WS$_2$ structure as an example [34], Figure 2 outlines two strategies: isovalent substitution and coupled substitution. For isovalent substitution, we first ranked potential replacement ions for $W^{4+}$ and $S^{2-}$ based on ionic similarity scores, and selected the top-10 candidates with the highest probabilities. Ions with identical oxidation states were then grouped into substitution sets. For example, candidate replacements for $W^{4+}$ included $Cr^{4+}$, $Mo^{4+}$, $Pd^{4+}$, and $Pt^{4+}$, while those for $S^{2-}$ included $Se^{2-}$, $O^{2-}$, and $Ge^{2-}$. By permutating combinations of these ions, a rich pool of isovalent substitution candidates was generated for further screening. For coupled substitution, we collected high-probability replacement ions for all atomic sites within a given prototype and classified them by oxidation state (such as +4, +3, -2, or -1). While maintaining overall charge neutrality (such as $Mo^{4+} + As^{3+} + 3S^{2-} + Br^{1-} = 0$), we carried out systematic substitutions across all structural sites using a combinatorial Cartesian product approach. This resulted in a diverse set of variable-valence candidate structures. To further enhance structural diversity, we applied supercell expansion techniques to enable fractional substitution and generate additional candidate configurations. In each iteration of the workflow, approximately $10^6$ structures were randomly sampled from the hundreds of millions of generated candidates for subsequent screening.

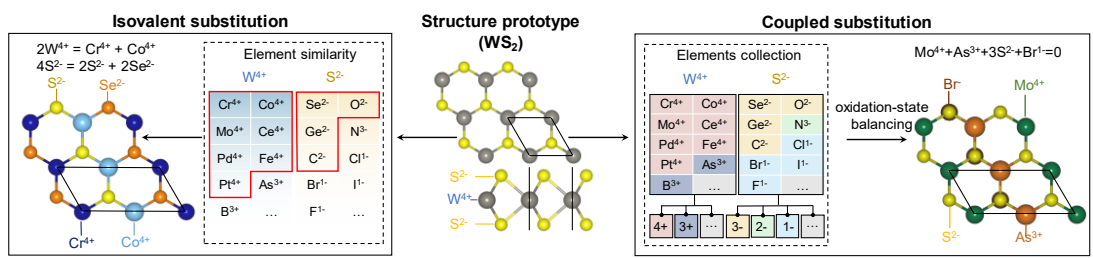

**Figure 2. Structural extension based on ionic similarity.** The structure expansion process involves isovalent and coupled substitution, which is guided by ionic similarity probabilities. For isovalent substitution, candidate ions with the same oxidation state as the original species in the structure are grouped into substitution sets. By enumerating distinct combinations within these sets, new candidate structures are generated for further screening. For coupled substitution, candidate ions for all substitutable sites in a given structural prototype are first categorized by their oxidation states. New candidate structures are then constructed by selecting combinations of these ions that satisfy overall oxidation-state balance, thereby enabling the exploration of variable-valence configurations.

**MLIP-accelerated structure screening**

Efficient structure optimization and formation energy estimation are essential for accelerating candidate material screening within the active learning workflow [35]. To this end, we modified the CHGNet model [36] under the Paddle framework [37] by redesigning its original graph batching mechanism. Though replacing the serial graph-to-batch approach (see the Methods section for more details), the computational structure screening efficiency is significantly enhanced. In each iteration, structural and energetic data obtained from DFT calculations in the previous round were incorporated into training the $CHGNet_{Paddle}$ model. The trained model was then used to optimize the geometry and predict the energy of new candidate structures. To improve model robustness, a subset of the MLIP-optimized structures ($Struc_{MLIP}^{opt}$) was further evaluated using single-point self-consistent DFT calculations. The corresponding energies were then added to the training set for subsequent iterations (see the Methods section for more details). To enable rapid thermodynamic stability assessment of the candidate structures, the convex hull energy ($E_{hull}^{MLIP}$) was calculated based on MLIP-predicted total energies. Decomposition phases along the convex hull pathway were sourced from both the OQMD database and the 2D materials dataset generated in this study. All decomposition energies ($E_{Decomp}^{DFT}$) were consistently obtained using DFT calculations at DFT-level accuracy.

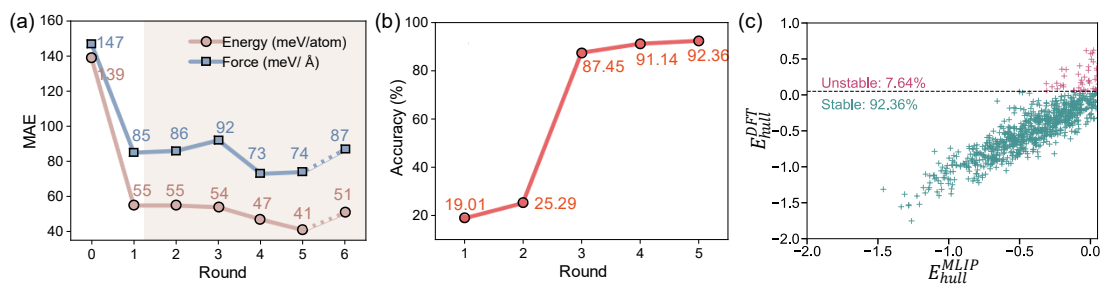

**Figure 3. Evolution of MLIP model performance.** (a) Energy and force prediction MAEs of the MLIP model of 2D materials decreased to 41 meV/atom and 74 meV/Å after five active learning iterations. The model achieved MAEs of 51 meV/atom for energy prediction and 87 meV/Å for force predictions when trained on all 242,546 newly generated structures. (b) Prediction accuracy for the identification of stable structures ($E_{hull}^{DFT}$) improves from 19.01% to 92.36% after 5 iterations. (c) For the 5[th] round of active learning, the thermodynamic stability predicted by the MLIP model ($E_{hull}^{MLIP}$) exhibits strong linear correlation with DFT-calculated values ($E_{hull}^{DFT}$).

As shown in Figure 3a, the CHGNet$_{Paddle}$ model trained on 21,684 2D materials achieved an initial MAE of 139 meV/atom for energy and 147 meV/Å for atomic forces. As the number of active learning iterations increased, both energy and force prediction errors exhibited a clear downward trend. After 5 rounds, the obtained energy prediction MAE and force prediction MAE trained on a dataset containing 1,024,059 structure–energy–force mappings derived from 207,106 2D materials decreased to 41 meV/atom and 74 meV/Å, respectively. Given the strong predictive performance of the CHGNet$_{Paddle}$ model at this stage, the model obtained from the 5[th] iteration was directly used for energy predictions in subsequent rounds. However, increasing the amount of training data beyond this point did not yield further improvements in the accuracy of energy and force predictions. The final iteration trained on a dataset containing

1,863,788 structure–energy–force mappings derived from 392,319 2D materials outputs an energy prediction MAE of 51 meV/atom and a force prediction MAE of 87 meV/Å. This observation suggests that the current model capacity or data diversity may be limiting factors and warrants further investigation.

**DFT-based validation**

To ensure the quality and reliability of the expanded 2D materials dataset, all candidate structures with $E_{hull}^{MLIP}$ < 50 meV/atom selected by MLIP in each iteration were subjected to DFT validation. High-throughput calculations were performed using JAMIP [38] to re-evaluate the $E_{hull}^{DFT}$ of these structures, providing a more precise assessment of their thermodynamic stability. Stable structures with $E_{hull}^{DFT}$ < 50 meV/atom were incorporated into the materials database and used as inputs for the next round of active learning. On the other hand, intermediate structures from both high-accuracy DFT optimization and lower-accuracy CHGNet$_{Paddle}$ optimization processes were uniformly sampled. Their DFT total energies were computed and used to further refine the CHGNet$_{Paddle}$ model in the subsequent training cycle.

As shown in Figure 3b, active learning workflow led to a substantial improvement in the CHGNet$_{Paddle}$ model's ability to identify stable structures with $E_{hull}^{MLIP}$ < 50 meV/atom. During the first three iterations, the prediction accuracy increased rapidly from 19.01% to 87.45%. After the 5$^{th}$ round, it reached the highest accuracy of 92.36%. Figure 3c presents a comparison between model predicted thermodynamic stabilities ($E_{hull}^{MLIP}$) and those DFT calculated results ($E_{hull}^{DFT}$). A strong linear correlation is observed between them, with only 7.64% of the structures incorrectly classified as unstable. It is the first generic MLIP for 2D materials trained over the periodic table with excellent optimization capability for unknown structures, which is partially validated in the subsequent structural energy prediction of diffusion models. These results highlight the robustness and generalization capacity of the CHGNet$_{Paddle}$ model after multiple rounds of active learning, enabling efficient and accurate identification of novel stable 2D materials.

**Dataset of 2D materials**

Building upon the active learning workflow described above, we developed the

ML2DDB, a comprehensive database containing over 242,546 DFT-validated monolayer structures with thermodynamic stability characterized by $E_{hull}^{DFT} < 50$ meV/atom. Compared to similar 2D datasets, the ML2DDB represents at least an order-of-magnitude increase in the total number of entries. As shown in Figure 4a, the dataset exhibits a wide elemental distribution, spanning 81 elements and covering nearly the entire periodic table except for radioactive and noble gas elements. Figure 4b illustrates the distribution of elemental diversity within the material structures. Compared with existing datasets, our collection shows substantial gains in the number of compounds containing three or four distinct elements, a category that has been challenging to discover using previous approaches. Representative examples are displayed in Figure 4c, encompassing a range of structural prototypes and diverse cation–anion combinations. These results highlight both the structural diversity of the dataset and the effectiveness of the proposed expansion strategy. Additionally, this process also yielded a larger dataset of over one million 2D structures with $E_{hull}^{MLIP} < 200$ meV/atom, offering valuable resources for future investigations into emerging 2D materials.

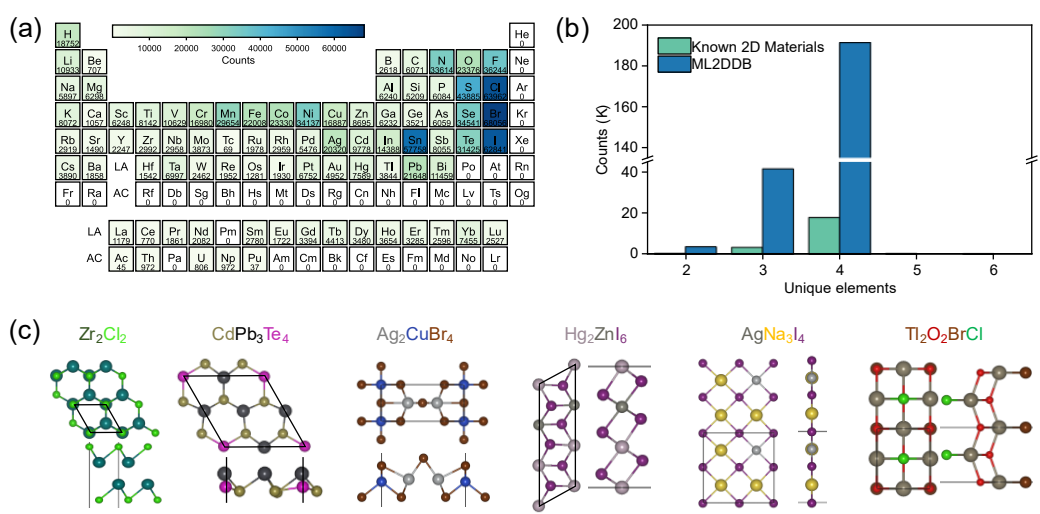

**Figure 4. Overview of ML2DDB.** (a) Elemental distribution heatmap of ML2DDB, covering 81 elements. (b) The number of ternary and quaternary structures shows a substantial increase compared to existing 2D materials datasets. (c) Representative examples of newly discovered stable 2D materials.

**Diffusion model generation of S.U.N. materials**

Samples are generated by inverting a fixed diffusion model of the damage process using a learned fractional network. Gaussian noise is usually added to the image of the damage process [39,40], and customized diffusion processes are needed because 2D structure always has a unique periodic structure and symmetry [41]. We implement the joint diffusion of element type (*A*), coordinate (*X*) and periodic lattice (*L*) based on the Paddle framework in MatterGen [42]. Specifically, the Normal distribution for coordinate diffusion using packing follows periodic boundaries and approaches a uniform distribution at the noise limit. The effect of unit cell size on the diffusion of fractional coordinates in Cartesian space is adjusted by correspondingly scaling the noise amplitude. Lattice diffusion is implemented in a symmetric form and is centered on a distribution of cubic lattices whose mean atomic density is taken from the training data. Atomic species are diffused in a categorical space, in which individual atoms are corrupted into a masked state. To reverse the corruption process, a score network was trained to output invariant scores for atomic species and equivariant scores for both coordinates and lattice parameters, without any requirement to learn symmetry from the data. Simultaneously, building on our existing framework, the specific thickness of 2D materials is introduced into the model as a vectorial embedding (see the Methods section for more details), whereby the diffusion model can be trained efficiently and can generate plausible structures during both the training and sampling processes. And an adapter module is introduced, through which the generation of 2D materials is guided along directions constrained by the target properties. In Figure 5a, we displayed a few random samples generated by the diffusion model, all of which have distinct 2D material features with reasonable coordination environments.

The capability to generate S.U.N. (stable, unique, new) 2D materials are prerequisites for diffusion models [43–46]. We considered a generated structure as stable with $E_{hull}^{DFT}$ < 100 meV/atom with respect to ML2DDB. The unique is specified whether a generated structure matches any other structure generated in the same batch or not, and the new is whether it is identical to any of the structures in ML2DDB. As shown in Figure 5b, we performed DFT structure optimization on 1024 structures to evaluate the stable attribute. The results show that 74.8% of them are considered stable ($E_{hull}^{DFT}$ < 100 meV/atom), which is comparable to the success rate of 3D stable structure generation of MatterGen [42]. When the constraint is set to $E_{hull}^{DFT}$ < 0 meV/atom, our method achieved a success rate of 59.6%, which is significantly higher than that of

MatterGen (~13%). In addition, the Root-mean-square displacement (RMSD) of the generated structure is lower than 0.26 Å compared to the DFT relaxation structure, which is still less than the radius of the hydrogen atom (0.53 Å) [47]. For the generation of unique structures, the success rate accounts for 100% when generating one thousand structures. The rate only decreases 4.4% when generating ten thousand structures. For the generation of new structures, the rate decreases from 100% to 73.5% when the generated structures grow from one thousand to two thousand. This indicates that our model has a relatively excellent ability to generate completely new stable structures.

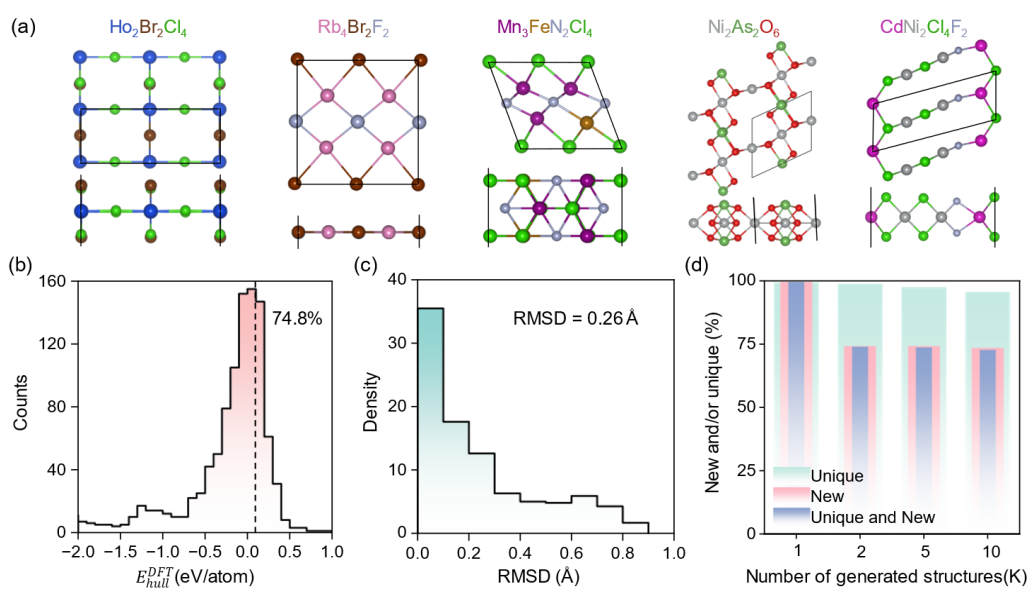

**Figure 5. Generation of stable, unique and new 2D materials.** (a) Visualization of five randomly selected crystals generated with corresponding chemical formula. (b) In the set of 1,024 structures generated via diffusion, 74.8% were confirmed as thermodynamically stable ($E_{hull}^{DFT} < 100$ meV/atom). (c) RMSD distribution between the initial generated structure and the DFT-relaxed structure. (d) Percentage of unique, new structures as a function of the number of generated structures.

**Chemistry system guided phase diagram construction**

Combining the diffusion-based generation model with high precision MLIP, we carried out stable structure search of 2D materials for different chemical systems. Compared with traditional crystal structure search methods, which often require tens or even hundreds of thousands of calculations to obtain a few candidate structures [48] for a single system, the present method is significantly accurate and efficient. Taking V-Se-O as an example (Figure 6a), we use the trained MLIP to make rapid stability predictions. The MLIP predicted ternary phase diagram is consistent with DFT verified ternary phase diagram (Figure 6b and 6c). The corresponding MAE for the MLIP predicted and DFT validated energy is only 54 meV/atom (Figure 6d), demonstrating the excellent optimization capability for unknown structures of our proposed MLIP. In V-Se-O system, we identified a variety of novel 2D crystal structures on convex hull (Figure 6e), among which $V_2Se_2O$ has been previously reported [49–51]. This indicates that the diffusion generation model not only recapitulates the known structures, but also accurately focuses on the thermodynamically stable 2D structures.

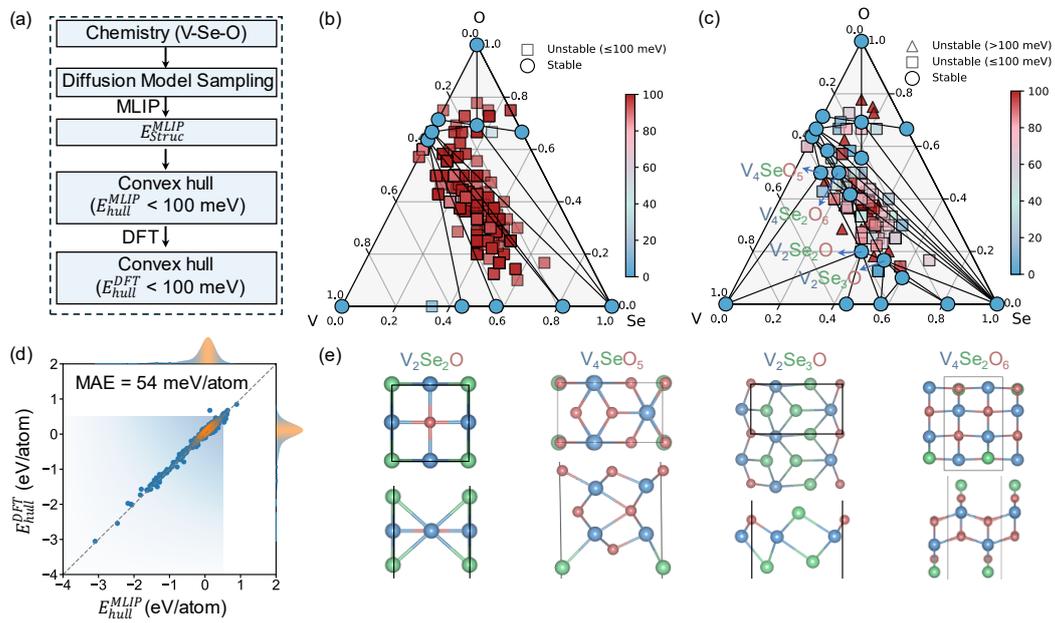

**Figure 6. Generation of materials in the target chemical system.** (a) The workflow of material generation, energy prediction, and convex hull construction during phase diagram construction. (b) Convex hull diagram plotted by MLIP prediction energies for the V-Se-O ternary system. (c) Convex hull diagram plotted by DFT validated energies for the V-Se-O ternary system. The stable structure is colored as blue circles, and the stability of metastable structures is visually encoded using colored boxes, which denote their energy distance above the convex hull. (d) The MAE for the MLIP predicted and DFT validated energy is only 54 meV/atom. (e) Four stable structures found in V-Se-O system after DFT validation.

**Space group constrained structure generation**

The spatial symmetry of the crystal not only determines the electronic energy bands and phonon vibrational properties [52,53], but also plays a decisive role in the existence and strength of second-order nonlinear optical responses [54] (e.g., second harmonic generation, SHG). Since only crystals without spatial inversion centers can have a non-zero second-order polarization tensor χ2, which generates a polarization component with a frequency of 2ω [55]. Any space group containing an inversion center is strictly 'forbidden' by symmetry to produce an effective SHG response. It has been a challenging task to accurately construct SHG-active materials with targeted noncentrosymmetric structures on the atomic scale without relying on a priori constraints on the symmetry of known materials [56–58]. The underlying generative model is fine-tuned by introducing spatial group labels to enhance its ability to generate specific noncentrosymmetric structures. As shown in Figure 7a, we generated 3200 candidate structures for each of the three typical SHG space groups *Pm*, *Pmm*2 and *P*3*m*1 to validate the ability of the model in enhancing the generation of target symmetry structures. The results show that the attribution ratios of the generated structures in all three space groups are more than 25%, which is significantly higher than the original ML2DDB training dataset. Among them, the generation ratio of the *Pm* space group is 10 times higher than that of the training set. Figure 7b shows the configurations of some of the typical generating structures under each space group. This result demonstrates that the fine-tuning strategy based on space group labelling can effectively guide the model to focus on the target symmetry and significantly improve the accuracy of generating 2D noncentrosymmetric materials.

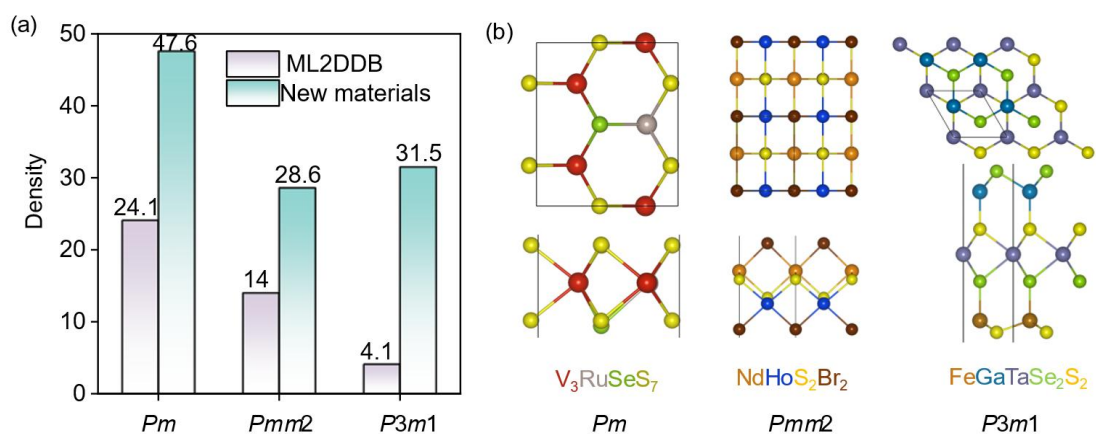

**Figure 7. Generation of materials with target symmetry**. (a) Comparison of the proportion of the three non-centrosymmetric space group structures generated *Pm*, *Pmm*2, *P*3*m*1 with the space group distribution of ML2DDB. (b) Random selection of three 2D structures generated with given space group.

## Conclusion

By combining an active learning workflow and conditional diffusion-based structure generation, our work achieves a significant expansion in the scale of 2D materials data and facilitates the generation of novel structures defined by specified elemental components or properties. The proposed ML2DDB exceeds at least an order-of-magnitude compared to existing datasets. Eventually, over 242,546 novel and thermodynamically stable 2D materials with $E_{hull}^{DFT}$ < 50 meV/atom were identified. The number of ternary compounds increased by 1100% and the number of quaternary compounds by 960%, thereby significantly enhancing the chemical diversity of the generated structures. Additionally, more than one million candidate structures with $E_{hull}^{MLIP}$ < 200 meV/atom were generated, greatly broadening the landscape for 2D materials discovery. The MLIP model trained on this dataset demonstrated strong predictive capability for stability classification, achieving an accuracy of 92.36%. As the diffusion models are introduced into the module, fine-tuning of the property labels enables the generation of phase diagrams for arbitrary chemical ratios as well as the generation of specified space group structures. This not only provides an intuitive analysis for thermodynamic stability analysis of 2D monolayer materials, but also offers the possibility of predicting new materials in the field of materials such as nonlinear optically responsive materials and ferroelectric materials. We anticipate that our workflow can be extended to other material properties, including carrier mobility, band gap, and magnetism.

Despite these advances, we recognize that several key challenges remain in bridging the gap between theoretical discovery and experimental synthesis of 2D materials. These include the understanding of phase transition mechanisms among competing polymorphs, the combined consideration of dynamical stability and configurational entropy, and the final prediction of synthesizability, all of which require further in-depth investigation.

# Methods
## Candidate structure generation via ionic substitution

We obtain candidate structures using a probabilistic model based on data-mined ion substitution probabilities [19]. Guided by these ionic similarity scores, atomic positions in each structural prototype are replaced in descending order of ion substitution probability. Specifically, the ionic substitution probability is defined as:

$$p(X, X') \approx \frac{\exp \sum_i \lambda_i f_i^{(n)}(X, X')}{Z}$$

Where $X$ and $X'$ represent vectors composed of $n$ distinct ions. The function $f_i$ is defined as 1 when a specific substitution pair occurs, and 0 otherwise. $\lambda_i$ denotes the weight assigned to of a given substitution and $Z$ is a partition function ensuring the normalization of the probability.

In this study, we refined the original probabilistic model to enlarge the candidate materials space and prioritize the discovery of previously unexplored compounds. The original formulation of conditional probabilities was inherently biased toward frequently observed substitution pairs in existing datasets. To mitigate this and promote the inclusion of rare combinations, we modified the model by uniformly setting the minimum substitution probability to zero. Starting from known compositions, we applied the physics-guided substitution probabilities to identify plausible candidate ions. Partial substitutions were then performed using a Cartesian product over all relevant atomic sites, ensuring comprehensive enumeration of inequivalent configurations and yielding a diverse dataset for subsequent screening.

## MLIP model

We adopted the CHGNet model for the structure optimization of 2D materials. This model encodes interatomic interactions using two distinct graph representations: the atom graph and the bond graph. Through a message-passing mechanism, it iteratively updates atomic, bond, and angular features to predict key material properties such as total energy and atomic forces. In the atom graph, nodes correspond to atomic numbers $Z_i$, and edges represent interatomic distances $r_{ij}$. The Bond Graph is constructed by treating edges in the Atom Graph as nodes, where edges between them denote the angles $\theta_{ijk}$ formed between two connected bonds. Following the construction of these two graphs, embeddings for the nodes and edges of both graphs

are generated as network features:

$$v_i^0 = Z_i W_v + b_v,$$

$$e_{ij,n}^0 = \tilde{e}_{ij} W_e, \tilde{e}_{ij} = \sqrt{\frac{2}{5}} \frac{\sin\left(\frac{n\pi r_{ij}}{5}\right)}{r_{ij}} \odot u(r_{ij}),$$

$$a_{ijk,\ell}^0 = \begin{cases} \frac{1}{\sqrt{2\pi}} & \text{if } \ell = 0 \\ \frac{1}{\sqrt{\pi}} \cos\left[\ell \theta_{ijk}\right] & \text{if } \ell = [1, N] \\ \frac{1}{\sqrt{\pi}} \sin\left[(\ell - N)\theta_{ijk}\right] & \text{if } \ell = [N+1, 2N] \end{cases}.$$

Where $W$ and $b$ are trainable parameters, $u(r_{ij})$ s the polynomial envelope function, subscript $n, \ell$ is the expansion orders. $\odot$ is the element-wise product. Where:

$$\theta_{ijk} = \arccos \frac{e_{ij} \cdot e_{jk}}{|e_{ij}||e_{jk}|},$$

the message passing policy in the CHGNet model is:

$$v_i^{t+1} = v_i^t + L_v^t \left[\sum_j \tilde{e}_{ij} \cdot \phi_v^t(v_i^t \| v_j^t \| e_{ij}^t)\right],$$

$$e_{jk}^{t+1} = e_{jk}^t + L_v^t \left[\sum_i \tilde{e}_{ij} \cdot \tilde{e}_{jk} \cdot \phi_e(e_{ij}^t \| e_{jk}^t \| a_{ijk}^t \| v_j^{t+1})\right],$$

$$a_{ijk,f}^{t+1} = a_{ijk}^t + \phi_a^t(e_{ij}^{t+1} \| e_{jk}^{t+1} \| a_{ijk}^t \| v_j^{t+1}).$$

where L is a linear layer and $\phi$ is a gated MLP

$$L(x) = xW + b,$$
$$\phi(x) = (\sigma \circ L_{\text{gate}}(x)) \odot (g L_{\text{core}}(x)),$$

$\sigma$ and $g$ are the Sigmoid and SiLU activation functions, respectively.

The energy is calculated by the nonlinear projection of the point-by-point averaged feature vectors on all atoms, and the force is calculated by self-differentiation of the energy with respect to the Cartesian coordinates of the atoms:

$$E_{\text{tot}} = \sum_i L_3 \circ g \circ L_2 \circ g \circ L_1(v_i^4),$$

$$\vec{f_i} = -\frac{\partial E_{\text{tot}}}{\partial \vec{x_i}}.$$

In the original CHGNet framework (https://github.com/CederGroupHub/chgnet), the sequential processing mechanism was employed during model training for embedding feature computation on batches of graph data. This serial execution pattern resulted in suboptimal GPU resource utilization. To enhance computational efficiency, we propose an optimized parallelization scheme: through batch graph concatenation, all graph structures within a single batch are tensor-concatenated to enable simultaneous embedding feature extraction across all graph instances. This strategy significantly improves GPU parallel computing utilization. For training supervision, we adopt the Mean Squared Error (MSE) loss function to construct the optimization objective:

$$\mathcal{L}(x, \hat{x}) = \frac{1}{N} \|x - \hat{x}\|_2^2,$$

where N is the number of samples. The loss function is the summary of energy and force：

$$\mathcal{L} = \mathcal{L}(E, \hat{E}) + \mathcal{L}(\mathbf{f}, \hat{\mathbf{f}}).$$

**DFT calculation**

To ensure computational consistency, a unified parameter set was implemented based on the plane-wave pseudopotential approach within density functional theory. The Perdew-Burke-Ernzerhof (PBE) exchange-correlation functional [59] was employed within the Vienna ab initio simulation package (VASP) [60]. Electron-ion interactions were described using the projector augmented wave (PAW) pseudopotentials [61]. Structural optimization (including lattice parameters and internal atomic positions) was performed using the conjugate gradient algorithm with a convergence threshold for residual forces below 0.02 eV/Å. A kinetic energy cutoff of 520 eV was applied for plane-wave expansion of electronic wavefunctions. Brillouin zone integration utilized a Monkhorst−Pack [62] $k$ mesh of $2\pi \times 0.03$ Å$^{-1}$ and the value along vacuum layer direction is set to 1. Long-range van der Waals interactions between layers were accounted for by the vdW-optB88 functional [63] to accurately describe weak interlayer and out-of-plane interactions in 2D materials. Electronic correlation effects were improved using the GGA+U approach for the exchange-correlation potential [64], where the effective on-site Coulomb interaction strength was applied. High-throughput DFT calculations were executed using the Jilin Artificial-intelligence aided Materials-design Integrated Package (JAMIP) – an open-source, AI-aided data-driven infrastructure specifically designed for computational materials informatics [38].

**Diffusion model**

MatterGen [42] is a diffusion model [43–46] whose core principle operates as follows: during the training phase, controlled noise is introduced into crystal structure data, and the model is trained to reverse this noise injection (denoising). This process allows the network to learn the intrinsic patterns for recovering ordered structures from random perturbations. during the sampling phase, the model takes a randomly initialized structure as input and progressively optimizes atomic species and spatial arrangements through multi-step iterative denoising, ultimately converging to thermodynamically stable crystal configurations. This generative framework based on diffusion probabilistic models effectively simulates the structural evolution from disorder to order.

The structural representation of a crystal can be defined through its atomic species matrix, lattice, and fractional coordinates:

$$M = (A, X, L),$$

where $A = (a^1, a^2, ..., a^n)^T \in \mathbb{A}^n$ represents the atomic species within the unit cell; $X = (x^1, x^2, ..., x^n) \in [0,1)^{3 \times n}$ denotes the fractional coordinate matrix for corresponding atoms [65,66]; $L = (l^1, l^2, l^3) \in \mathbb{R}^{3 \times 3}$ corresponds to the lattice. For crystal structure diffusion and denoising processes, the operations can be systematically decomposed by separately applying noise perturbation and restoration to the three fundamental components: atomic species matrix ($A$), fractional coordinate matrix ($X$), and lattice constant matrix ($L$) of the crystal structure.

For discrete atomic types, MatterGen employs the D3PM [67] framework for diffusion-denoising modeling. Its forward diffusion process follows a Markov chain, achieving progressive structural disruption of input samples through stepwise perturbed discrete state transitions. In each diffusion step, the model randomly replaces atomic types based on a transition probability matrix, ultimately transforming the atomic types in the original crystal structure into completely random noise. The reverse denoising process learns the inverse mapping through a parameterized Markov chain, progressively reconstructing coherent atomic types. The forward diffusion process is defined as:

$$q(a_{1:T}|a_0) = \prod_{t=1}^{T} q(a_t|a_{t-1}),$$

Where $a_0 \sim q(a_0)$ represents atomic types sampled from the data distribution and

$a_T \sim q(a_T)$, where $q(a_T)$ denotes a prior distribution.

By encoding $a$ as a one-hot row vector $\boldsymbol{a}$, the transition probability at each diffusion step is defined as:

$$q(\boldsymbol{a}_t|\boldsymbol{a}_{t-1}) = Cat(\boldsymbol{a}_t; \boldsymbol{p} = \boldsymbol{a}_{t-1}\boldsymbol{Q}_t),$$

where $[\boldsymbol{Q}_t]_{ij} = q(a_t = j|a_{t-1} = i)$ represents the Markov transition matrix at time step $t$. $Cat(\boldsymbol{a}; \boldsymbol{p})$ denotes a categorical distribution over a one-hot vector with probabilities specified by the row vector $\boldsymbol{p}$.

In the MatterGen model, the Variance-exploding method [68] is employed for the diffusion and denoising processes of fractional coordinates. However, due to the strong correlation between atomic coordinates in Cartesian space and unit cell dimensions, conventional approaches that add noise to fractional coordinates using fixed variance strategies exhibit significant limitations. To overcome this bottleneck, MatterGen innovatively proposes a dynamic variance adjustment mechanism based on atomic density distribution. This method abandons the traditional fixed constant variance paradigm and instead constructs a variance modulation strategy tailored to atomic density distribution characteristics, enabling adaptive optimization of noise injection intensity. The calculation formula is as follows:

$$\sigma_t(n) = \frac{\sigma_t}{\sqrt[3]{n}},$$

where $\sigma_t$ represents the original variance at time step $t$, and $n$ denotes the number of atoms within the unit cell.

MatterGen employs a variance-preserving approach for diffusion and denoising of lattice constants. To achieve rotational invariance in material structures, the method utilizes singular value decomposition (SVD)-based polar decomposition to transform the lattice into a symmetric positive definite (SPD) matrix, followed by performing diffusion and denoising operations on this symmetric matrix.

The decomposition follows the matrix equations:

$$\tilde{\boldsymbol{L}} = \boldsymbol{U}\boldsymbol{L}, \quad \boldsymbol{U} = \boldsymbol{W}\boldsymbol{V}^T, \quad \boldsymbol{L} = \boldsymbol{V}\boldsymbol{\Sigma}\boldsymbol{V}^T,$$

where $\boldsymbol{W}$ and $\boldsymbol{V}$ represent the left and right singular vectors of $\tilde{\boldsymbol{L}}$ respectively, and $\boldsymbol{\Sigma}$ is the diagonal matrix of singular values. $\boldsymbol{U}$ is a rotation matrix and $\boldsymbol{L}$ is a symmetric positive-definite matrix.

The following constitutes the loss function during model training, comprising two components: the score matching loss for coordinates and lattice constants, and the atomic type classification loss:

$$L = \lambda_{cord}L_{coord} + \lambda_{cell}L_{cell} + \lambda_{types}L_{types},$$

where:

$$L_{coord} = \sum_{t=1}^{T} \sigma_t(n)^2 \mathbb{E}_{q(x_0)} \mathbb{E}_{q(x_t|x_0)} \left[ \left\| s_{x,\theta}(X_t, L_t, A_t, t) - \nabla_{x_t} \log q(x_t|x_0) \right\|_2^2 \right],$$

$$L_{cell} = \sum_{t=1}^{T} (1 - \bar{\alpha}_t)\sigma_t(n)^2 \mathbb{E}_{q(L_0)} \mathbb{E}_{q(L_t|L_0)} \left[ \left\| s_{L,\theta}(X_t, L_t, A_t, t) - \nabla_{L_t} \log q(L_t|L_0) \right\|_2^2 \right],$$

$$L_{types} = \mathbb{E}_{q(a_0)} \left[ \sum_{t=2}^{T} \mathbb{E}_{q(a_t|a_0)} [D_{KL}[q(a_{t-1}|a_t, a_0) || p_\theta(X_t, L_t, A_t)] \right.$$

$$\left. - \lambda_{CE} \log p_\theta(a_0|X_t, L_t, A_t, t)] - \mathbb{E}_{q(a_1|a_0)}[\log p_\theta(a_0|X_1, L_1, A_1, 1)] \right],$$

where $L_{coord}$ and $L_{types}$ show the loss only for a single atom's coordinates and specie, respectively; the overall losses for coordinates and atom types sum over all atoms in a structure.

The primary objective of the MatterGen network model is to predict crystal structure scores, including atomic types, atomic positions, and lattice. We will first elaborate on how MatterGen predicts these three components, followed by an introduction to the architectural components of the MatterGen network: Graph Construction, Equivariant Scoring Network, and Adapter Module. Additionally, in response to the characteristic limitation of spread along the z-axis in 2D materials, we have incorporated a dimensional vector into the MatterGen framework.

During the denoising process, MatterGen employs an SE(3)-equivariant Graph Neural Network (GNN) to predict scores for atomic positions, atomic types, and lattice. For atomic coordinates, MatterGen first predicts Cartesian coordinate scores $s_{X,\theta}(X_t, L_t, A_t, t)$, which are then converted into fractional scores using the following formula:

$$X = L^{-1}\widetilde{X},$$

where $X$ represents fractional coordinates, $\widetilde{X}$ denotes Cartesian coordinates, and $L$ corresponds to the lattice.

For atomic type prediction, MatterGen estimates the atomic species $A_0$ at the initial timestep $t = 0$ based on the output of the final message-passing layer in the GNN model. The input to this prediction module consists of the crystal structure information $(X_t, L_t, A_t, t)$ at timestep $t$, formulated as:

$$\log p_\theta(A_0|X_t, L_t, A_t, t) = H^{(L)}W,$$

where $H^{(L)} \in \mathbb{R}^{n \times d}$ denotes the output features from the last message-passing layer of the GNN, $W \in \mathbb{R}^{d \times K}$ represents the weights of the fully connected linear layer, and $K$ corresponds to the total number of atomic species (including masked null states).

For lattice scores, MatterGen incorporates rotational equivariance and scale invariance properties through Cartesian coordinate matrix operations and normalization. The model computes lattice scores at each GNN layer and aggregates results across all layers:

$$\widetilde{\Phi}^l = diag\left(\frac{\phi^l(m_{ijk}^l)}{|\varepsilon| \cdot d_{ijk}^2}\right),$$
$$s_{L,\theta}^l(X_t, L_t, A_t, t) = \widetilde{D}\widetilde{\Phi}^l\widetilde{D}^T,$$
$$s_{L,\theta}(X_t, L_t, A_t, t) = \sum_{l=1}^{L} s_{L,\theta}^l(X_t, L_t, A_t, t),$$

where $m_{ijk}^l \in \mathbb{R}^d$ denotes the edge features between atom $i$ (in the central unit cell) and atom $j$ (displaced by $k \in \mathbb{Z}^3$ unit cells) at layer $l$, $\phi^l$ represents a multi-layer perceptron (MLP), $d_{ijk}$ is the Euclidean distance between atoms $i$ and $j$ in fractional coordinates, $|\varepsilon|$ denotes the total number of edges, $\widetilde{D} \in \mathbb{R}^{3 \times |\varepsilon|}$ is the stacked matrix of Cartesian distance vectors.

To address the periodicity inherent in crystalline systems, MatterGen employs a directed multi-graph $G = (V, E)$ to represent each crystal structure, where $V = \{v_i\}_{i=1:N^v}$ denotes the nodes of the graph, and each node $v_i$ represents the feature vector of atom $i$ in the crystal structure. $E = \{e_{ij,(k_1,k_2,k_3)} | i, j \in \{1, ..., N\}, k_1, k_2, k_3 \in \mathbb{Z}\}$ denote the edges of the graph, where $e_{ij,(k_1,k_2,k_3)}$ denotes a directed edge pointing to node $j$ in the cell from the node $i$ in the original cell to the transvector $k_1 l_1 + k_2 l_2 + k_3 l_3$ a directed edge of node $j$ in the shifted cell.

MatterGen employs the GemNet architecture to predict scores for atomic positions, atomic types, and lattice during the denoising process. Originally developed as a general-purpose machine learning force field (MLFF), GemNet is a symmetry-aware message-passing graph neural network (GNN) that achieves SO(3)-equivariance through directional message passing [69]. The architecture enhances computational efficiency by integrating two- and three-body information within the initial network layers. Since energy prediction is not required, MatterGen utilizes the direct force prediction variant of this architecture—GemNet-dT.

To enable controllable crystal generation under property constraints, MatterGen integrates an Adapter Module into the unconditional scoring network for fine-tuning [70]. This adapter incorporates property information into the GemNet scoring architecture through an embedding layer and multi-layer adapters: an embedding layer $f_{embed}$ generates property vectors g from input constraints; Four adapter layers $f_{adapter}^{(L)}$ (two-layer MLPs) are inserted before each message-passing layer. A zero-initialized mix-in layer [71] $f_{mixin}^{(L)}$ dynamically combines property features with original node representations $H_j^{(L)}$ :

$$H_j'^{(L)} = H_j^{(L)} + f_{mixin}^{(L)}\left(f_{adapter}^{(L)}(g)\right) \cdot \mathbb{I}(property\ is\ not\ null).$$

Gated Conditioning: Property-aware features are injected only when valid property labels are provided, implemented through an indicator function $\mathbb{I}()$. During fine-tuning, all parameters (original GemNet and new embedding/adapter/mix-in layers) are jointly optimized to enable effective coordination between input property constraints and the model's inherent geometric features

To address the monolayer thickness along the z-axis in 2D crystalline materials, we innovatively incorporate this structural information into the graph neural network architecture of the MatterGen model. During the training process of the 2D materials database, we explicitly encode the z-axis expansion characteristics into the crystal graph node features, enabling precise modeling of the spatial configuration of 2D materials.

The dimensional vector is defined as:

$$\boldsymbol{d}_{vec} = Abs(\boldsymbol{X}_z - 0.5),$$

the updated node initialization is formulated as:

$$\boldsymbol{H}'^{(0)} = \boldsymbol{H}^{(0)} + MLP\big(RBF(\boldsymbol{d}_{vec})\big),$$

where: $\boldsymbol{X}_z \in [0,1)^{1 \times n}$ represents the z components of atomic fractional coordinates, $Abs$ denotes the absolute value function, $RBF$ represents the radial basis function, $MLP$ represents the multilayer perceptron, $\boldsymbol{H}^{(0)}$ and $\boldsymbol{H}'^{(0)}$ correspond to the initial node representations before and after incorporating the expansion information, respectively.